\documentclass[lettersize, journal]{IEEEtran}

\usepackage{cite}

\ifCLASSINFOpdf
\else
\fi

\usepackage{graphicx, subfigure}
\usepackage{amsmath,amssymb,amsfonts,algpseudocode}
\usepackage{upgreek}
\usepackage{optidef}
\usepackage{multirow}
\usepackage{array}
\usepackage{stfloats}
\usepackage{amsthm}
\usepackage{lipsum}
\usepackage{mathtools}
\usepackage{cuted}
\usepackage{algorithm}
\usepackage{blindtext}
\usepackage[font=small]{caption}
\usepackage{color}

\def\NoNumber#1{{\def\alglinenumber##1{}\State #1}\addtocounter{ALG@line}{-1}}

\begin{document}

\title{Aerial STAR-RIS Empowered MEC: A DRL Approach for Energy Minimization}

\author{Pyae Sone Aung,
        Loc X. Nguyen,
        Yan~Kyaw~Tun,~\IEEEmembership{Member,~IEEE,}
        Zhu Han,~\IEEEmembership{Fellow,~IEEE,}
        and~Choong~Seon~Hong,~\IEEEmembership{Fellow,~IEEE}
\thanks{Pyae Sone Aung, Loc X. Nguyen and Choong Seon Hong are with the
Department of Computer Science and Engineering, Kyung Hee University,
Yongin-si, Gyeonggi-do 17104, Rep. of Korea, e-mail:\{pyaesoneaung, xuanloc088,
cshong\}@khu.ac.kr.}
\thanks{Yan Kyaw Tun is with the Department of Electronic Systems, Aalborg University, A. C. Meyers Vænge 15, 2450 København, e-mail: ykt@es.aau.dk.}  
\thanks{Zhu Han is with the Electrical and Computer Engineering Department,
University of Houston, Houston, TX 77004, and the Department of Computer
Science and Engineering, Kyung Hee University, Yongin-si, Gyeonggi-do
17104, Rep. of Korea, email\{hanzhu22\}@gmail.com.}}

\maketitle
\begin{abstract}
Multi-access Edge Computing (MEC) addresses computational and battery limitations in devices by allowing them to offload computation tasks. To overcome the difficulties in establishing line-of-sight connections, integrating unmanned aerial vehicles (UAVs) has proven beneficial, offering enhanced data exchange, rapid deployment, and mobility. The utilization of reconfigurable intelligent surfaces (RIS), specifically simultaneously transmitting and reflecting RIS (STAR-RIS) technology, further extends coverage capabilities and introduces flexibility in MEC. This study explores the integration of UAV and STAR-RIS to facilitate communication between IoT devices and an MEC server. The formulated problem aims to minimize energy consumption for IoT devices and aerial STAR-RIS by jointly optimizing task offloading, aerial STAR-RIS trajectory, amplitude and phase shift coefficients, and transmit power. Given the non-convexity of the problem and the dynamic environment, solving it directly within a polynomial time frame is challenging. Therefore, deep reinforcement learning (DRL), particularly proximal policy optimization (PPO), is introduced for its sample efficiency and stability. Simulation results illustrate the effectiveness of the proposed system compared to benchmark schemes in the literature.
\end{abstract}


%
\IEEEpeerreviewmaketitle

\vspace{-0.1in}
\section{Introduction}\label{intro}
The widespread adoption of Internet of Things (IoTs) devices, including smart home gadgets, wearables, medical equipment, and household appliances, has generated various challenges, including network latency, bandwidth limitations, and insufficient computational resources for these low-power devices. A potential solution approach involves the utilization of multi-access edge computing (MEC), where IoT devices can offload some of the computation tasks to edge servers, providing localized processing power at the network edge. MEC enables delivering adaptive and effective services where low latency and high-performance computing are critical, such as in smart cities, industrial automation, and healthcare domains. Nevertheless, the task of establishing a line-of-sight (LoS) connection remains challenging in urban areas filled with an abundance of tall structures. To address this problem, unmanned aerial vehicle (UAV) is implemented with the benefit of its rapid deployment and mobility. In \cite{zhan2011wireless}, the authors considered the UAV as a relay to maximize the uplink communication rate. While the installation of a UAV as a relay or mobile base station demonstrates notable advancements, this approach entails a substantial energy consumption. Due to its limited energy resources, apart from flying, UAVs must prioritize energy conservation. To substantiate this issue, reconfigurable intelligent surface (RIS) is advantageous owing to its cost-effectiveness and energy efficiency, alongside offering improved signal quality at the receiver. An RIS is a planar surface composed of several passive reflecting elements,which can alter the phase of propagation of wireless signals and, hence, contribute to enhanced spectral efficiency. In \cite{huang2019reconfigurable}, the authors studied the energy-efficient system for RIS-assisted wireless network while satisfying the minimum data rate for the users. Still, the restriction of RIS solely focusing on reflecting limits the the transmitter and receiver required to be on the same side. Therefore, the research has shifted towards simultaneously transmitting and reflecting RIS (STAR-RIS) for solving the coverage limitation of conventional RIS, by providing $360^{\circ}$ coverage, which leads to a better degree of freedom (DoF). The authors in \cite{mu2021simultaneously} proposed the novel concept STAR-RIS aided wireless communications, which exhibits superior power efficiency in comparison to conventional RIS.

In \cite{zhai2022energy}, the authors examined the conventional UAV-RIS-assisted MEC network, while the authors in \cite{zhang2023resource} investigated the STAR-RIS aided MEC system. To date, there has been no existing studies that focus on the interactions between aerial STAR-RIS and MEC. By employing the advantages in rapid deployment and mobility of UAV combined with enhanced spectral efficiency and coverage improvement of STAR-RIS, in this paper, we propose an aerial STAR-RIS-assisted MEC system. Then, we formulate a problem to minimize energy consumption by jointly optimizing the trajectory of aerial STAR-RIS, offloading portion of the task, the amplitudes and phase shift coefficients for transmission and reflection modes, and power allocation, respectively. Since the problem is non-convex and contains mixed integers, it is challenging to solve in polynomial time. Considering the dynamic nature of the environment and unpredictable outcomes, we utilize deep reinforcement learning (DRL) as a competent solution approach. Among DRL techniques, we apply proximal policy optimization (PPO) method due to its sample efficiency, stability, and capability to handle both discrete and continuous actions \cite{schulman2017proximal}. Numerical results indicate the effectiveness of our proposed system compared to the benchmark schemes.

\vspace{-0.1in}
\section{System Model}\label{systemmodel}
\begin{figure}[t]
    \centering
	\includegraphics[width=\linewidth]{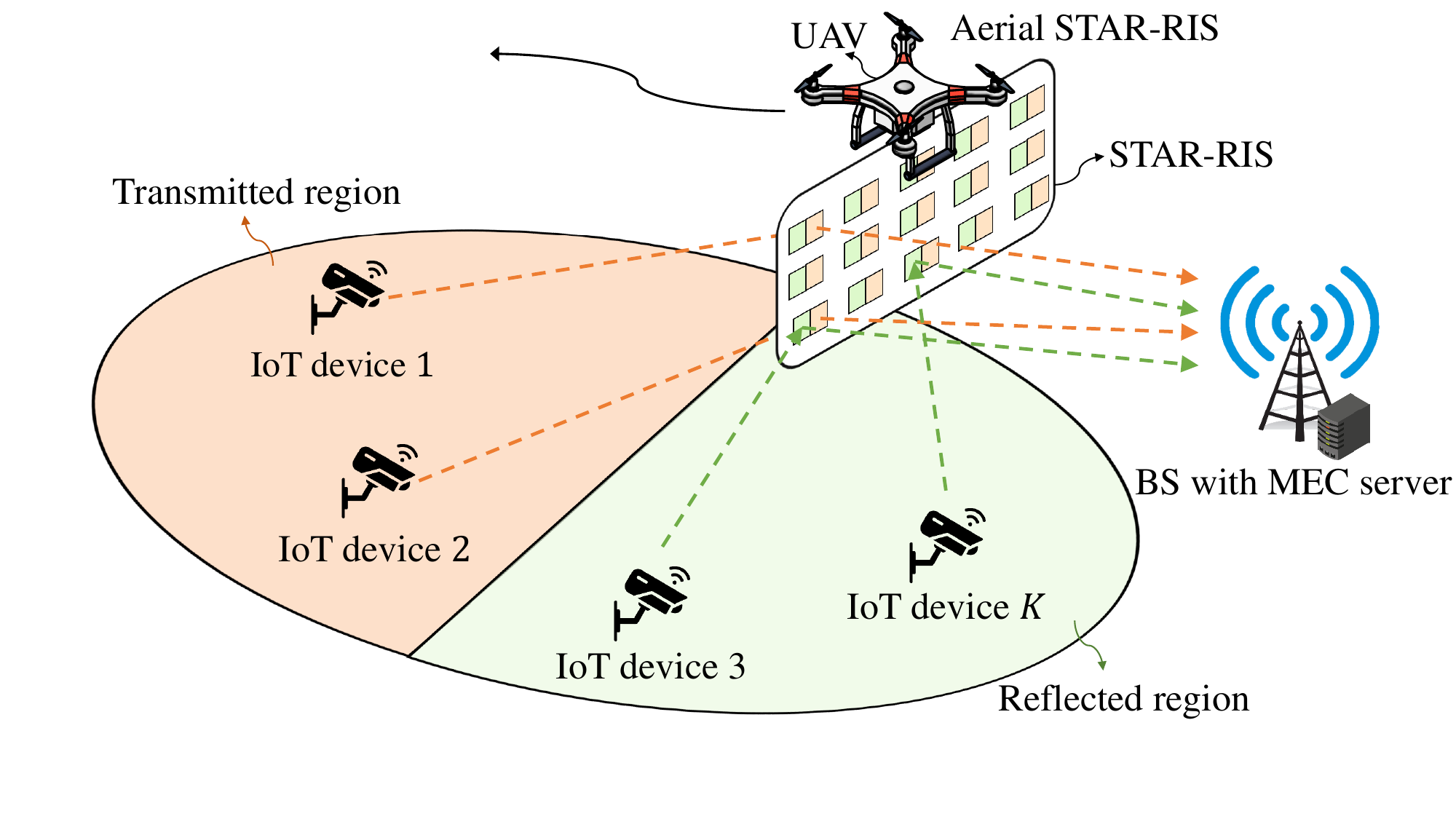}
	\caption{System model.}
	\label{sm}
\end{figure}
As illustrated in Fig.~\ref{sm}, with a single STAR-RIS implemented on a UAV, a base station (BS) integrated with an MEC server and multiple antennas $B$, and a set $\mathcal{K}$ of $K$ IoT devices, we consider an aerial STAR-RIS-assisted MEC system. Due to the obstacles, establishing line-of-sight (LoS) communication links between the BS and IoT devices can be a daunting or even unattainable task. Therefore, the aerial STAR-RIS is operated by the same service provider with the BS to assist the communication links and are built using low-latency wireless communication protocols like IEEE 802.11ax and IEEE 802.11be in order to decrease air-interface latency. The aerial STAR-RIS consists of an array of $\mathcal{M}=\{1,2, \dots, M\}$ elements, and each element contains reflection/transmission coefficients and phase shifters to direct the incident signal into the desired direction. With aerial STAR-RIS flying at the fixed altitude $H$, the positions of aerial STAR-RIS, BS, and IoT devices are denoted as $\boldsymbol{q} = \{x_s, y_s, H\}$, $\boldsymbol{v} = \{x_b, y_b, 0\}$, and $\boldsymbol{c} = \{x_k, y_k, 0\}$, respectively in the 3D Cartesian coordinate system. The IoT devices are divided into $R$ reflection IoT devices and $T$ transmission IoT devices, with $K = R + T$. The flight period for aerial STAR-RIS $T$ can be discretized into a set $\mathcal{N}$ of $N$ evenly spaced time intervals, each having a duration of $I=T/N$. Due to its power limitations, aerial STAR-RIS must return to its starting position at the end of the flight period, i.e., $\boldsymbol{q}(1)=\boldsymbol{q}(N)$ to recharge. Moreover, to guarantee that the speed of the aerial STAR-RIS does not exceed the maximum speed $V^{\max}$, we have the following constraint as
\begin{equation}
    \frac{||\boldsymbol{q}(n+1)-\boldsymbol{q}(n)||}{I} \leq V^{\max}, \forall{n} \in \mathcal{N}.
\end{equation}
Moreover, according to \cite{tun2020energy}, the energy consumption for the flight of aerial STAR-RIS ae each time slot can be obtained as
\begin{equation}
    E_{\mathrm{flight}}(n) = \kappa \left(\frac{||\boldsymbol{q}(n+1)-\boldsymbol{q}(n)||}{I}\right), \forall{n} \in \mathcal{N}, 
\end{equation}
where $\kappa = 0.5M$ and $M$ is the mass of aerial STAR-RIS, including UAV's payload.
\subsection{Local Computation Model}
For every IoT device $k$, a tuple of computation tasks denoted by $\{I_k(n), G_k,  T_k^{\max}(n)\}$ is assigned to be executed during each time slot. In this context, $I_k(n)$ denotes the size of the input data for each computation task, $G_k$ indicates the amount of required computing resources to process 1-bit of input data, and $T_k^{\max}(n)$ represents the maximum tolerable latency for the task completion. Due to the restricted energy and computing capability of IoT devices, it is not feasible to complete all tasks locally. As a result, particular tasks must be offloaded to the server. To enable partial offloading, we assume that the input task data bits are bit-wise independent, allowing them to be divided into subsets of any size and executed simultaneously by IoT devices and the MEC server. Therefore, we denote $\lambda_k(n) \in [0,1]$ as the portion of the task to offload to the MEC server, while $(1-\lambda_k(n))$ is the portion of the task to compute locally. With this, the execution latency for local computation for each IoT device $k$ at each time slot can be obtained as
\begin{equation}
    t_k^{\mathrm{local}}(n) = \frac{(1-\lambda_k(n))I_k(n) G_k}{f_k}, \forall{k} \in \mathcal{K}, \forall{n} \in \mathcal{N},
\end{equation}
where $f_k$ is the local computing capability of IoT device $k$. Next, the energy consumption for local computation of IoT device $k$ can be represented as
\begin{equation}
\resizebox{\hsize}{!}{$
    E_k^{\mathrm{local}}(n) = P_k^{\mathrm{local}}(n) t_k^{\mathrm{local}}(n) = \varrho f_k^2(1-\lambda_k)I_k(n) G_k, \forall{k} \in \mathcal{K}, \forall{n} \in \mathcal{N},
$}
\end{equation}
where $\varrho$ is the coefficient depending on the architecture of processor's chip.

\subsection{Communication Model}
To understand the operation of aerial STAR-RIS in our system, we denote $\mathbf{\Theta}^{\delta}(n) \in \mathbb{C}^{M \times M}, \delta \in \{r,t\}$ as the transmission and reflection coefficient matrices which can be specified by
\begin{equation}
\resizebox{\hsize}{!}{$
    \mathbf{\Theta}^r(n) = \operatorname{diag}(\sqrt{\beta^r_1(n)} e^{j{\phi}^r_1(n)}, \dots, \sqrt{\beta^r_M(n)} e^{j{\phi}^r_N(n)}), \forall{n} \in \mathcal{N},
$}
\end{equation}
\begin{equation}
\resizebox{\hsize}{!}{$
    \mathbf{\Theta}^t(n) = \operatorname{diag}(\sqrt{\beta^t_1(n)} e^{j{\phi}^t_1(n)}, \dots, \sqrt{\beta^t_M(n)} e^{j{\phi}^t_N(n)}), \forall{n} \in \mathcal{N},
$}
\end{equation}
where $\beta^r_m(n)$, $\beta^t_m(n) \in [0,1], \forall{m} \in \mathcal{M}$ are the amplitude coefficients of reflected and transmitted signal of $m$-th element, and ${\phi}^r_m(n)$, ${\phi}^t_m(n) \in [0, 2\pi), \forall{m} \in \mathcal{M}$ are phase shift values for reflection and transmission of the $m$-th element, respectively. We apply an energy-splitting model whereby the energy of the incoming signal on each element is conventionally partitioned into the energies of the transmitted and reflected signals \cite{mu2021simultaneously}. Therefore, with the law of conservation of energy, we have $\beta^r_n(n) + \beta^t_n(n) = 1, \forall{m} \in \mathcal{M}, \forall{n} \in \mathcal{N}$.

For the channel model, we assume there are no direct links between the IoT devices and BS due to the blockage by obstacles. The communication links are solely assisted by the aerial STAR-RIS. The channel gain from IoT device $r$ on the reflected region to the BS can be represented as
\begin{equation}
    h_r(n) = (\mathbf{h}_{M,B}(n))^H \mathbf{\Theta}^{r}(n) \mathbf{h}_{r,M}(n) \in \mathbb{C}^M, \forall{n} \in \mathcal{N},
\end{equation}
where $\mathbf{h}_{M,B}(n) \in \mathbb{C}^{M \times 1}$ is the channel response vector between the aerial STAR-RIS and BS, and $\mathbf{h}_{r,M}(n) \in \mathbb{C}^{M \times 1}$ is the channel response vector between IoT device in the reflected region $r$ to the aerial STAR-RIS, respectively. Similarly, the channel gain from IoT device $t$ on the transmitted region to the BS can be expressed as
\begin{equation}
    h_t(n) = (\mathbf{h}_{M,B}(n))^H \mathbf{\Theta}^{t}(n) \mathbf{h}_{t,M}(n) \in \mathbb{C}^M, \forall{n} \in \mathcal{N},
\end{equation}
where $\mathbf{h}_{t,M}(n) \in \mathbb{C}^{M \times 1}$ is the channel response vector between IoT device in the transmitted region $t$ to the aerial STAR-RIS. We assume the Rician channel model for  $\mathbf{h}_{M,B}(n)$,  $\mathbf{h}_{r,M}(n)$ and $\mathbf{h}_{t,M}(n)$, respectively.
For instance, $\mathbf{h}_{M,B}(n)$ can be expressed as
\begin{equation}
   \mathbf{h}_{M,B}(n) = \sqrt{\frac{\ddot{R}}{1+\ddot{R}}}\mathbf{h}_{M,B}^{\mathrm{LoS}}(n) + \sqrt{\frac{1}{1+\ddot{R}}}\mathbf{h}_{M,B}^{\mathrm{NLOS}}, \forall{n} \in \mathcal{N},
\end{equation}
where $\ddot{R}$ is the Rician factor, $\mathbf{h}_{M,B}^{\mathrm{LoS}}(n)$ is the deterministic LoS component, and $\mathbf{h}_{M,B}^{\mathrm{NLOS}}$ is the non-LoS component which is assumed to be independent and identically distributed circularly symmetric complex Gaussian random variable with zero-mean and unit variance.

Hence, the received signal at the BS can be obtained as
\begin{equation}
    y(n) = \sum_{k=1}^{K} h_k(n) x_k(n) + n_0, \forall{n} \in \mathcal{N},
\end{equation}
where $x_k(n) = s_k p_k(n)$ is the transmitted signal from IoT device $k$ with unit-power information symbol $s_k$, and transmit power $p_k(n)$ of IoT device $k$. The symbol $n_0$ is the additive white Gaussian noise (AWGN) with zero mean and variance $\sigma^2$. Consequently, the signal-to-interference-plus-noise ratio (SINR) of the IoT device at reflected region $r$ can be expressed as
\begin{equation}
    \gamma_r(n) = \frac{p_r(n) |h_r(n)|^2}{\sum_{j=r+1}^R p_j(n) |h_j(n)|^2 + \sigma^2}, \forall{n} \in \mathcal{N}.
\end{equation}

Likewise, the SINR of the IoT device at transmitted region $t$ can be represented as
\begin{equation}
    \gamma_t(n) = \frac{p_t(n) |h_t(n)|^2}{\sum_{j=t+1}^T p_j(n) |h_j(n)|^2 + \sigma^2}, \forall{n} \in \mathcal{N}.
\end{equation}
The wireless frequency of the BS is partitioned into orthogonal sub-carriers, each with a bandwidth denoted as $W$. Assuming that each BS-IoT device pair is on a different sub-carrier, the BS can decode the signal it receive on that sub-carrier, despite the STAR-RIS effectively reflecting or transmitting all incident signals. This means that there is no interference between BS-IoT device pairs \cite{aung2023deep}. The achievable data rate of IoT device $k$ can be obtained as
\begin{equation}
    r_k(n)= W \log_2\left( 1 + \gamma_\delta(n) \right), \delta \in \{r, t\}, \forall{n} \in \mathcal{N}. 
\end{equation}

Since the computational resources of IoT devices are limited and cannot perform all the tasks locally, a portion of the tasks must be offloaded to the MEC server. The time required for uplink transmission of offloaded portion of the task $\lambda_k(n)$ of IoT device $k$ can be represented as
\begin{equation}
    t_k^{\mathrm{off}}(n) = \frac{\lambda_k(n) I_k(n)}{r_k(n)}, \forall k \in \mathcal{K}, \forall{n} \in \mathcal{N}.
\end{equation}

Furthermore, the energy consumption for task offloading of IoT device $k$ can be expressed as
\begin{equation}
\resizebox{\hsize}{!}{$
E_k^{\mathrm{off}}(n) = P_k^{\mathrm{off}} (n) t_k^{\mathrm{off}}(n) = \frac{p_k(n) \lambda_k(n) I_k(n)}{r_k(n)}, \forall k \in \mathcal{K}, \forall{n} \in \mathcal{N}.
$}
\end{equation}

Since the computation power of the BS is high and the size of executed result is small enough, energy consumption, task execution time, and download latency from the MEC server are all disregarded \cite{bai2020latency}. Therefore, the total energy consumption for our proposed system can be obtained as
\begin{equation}
\resizebox{\hsize}{!}{$
E(\boldsymbol{\lambda}, \boldsymbol{q}, \boldsymbol{\phi}, \boldsymbol{\beta}, \boldsymbol{p}) = \left(\sum_{n=1}^{N} \sum_{k=1}^{K}  E_k^{\mathrm{local}}(n) + E_k^{\mathrm{off}}(n) \right) + \sum_{n=1}^{N}E_{\mathrm{flight}}(n).
$}
\end{equation}

\vspace{-0.2in}
\section{Problem Formulation}
Given our system model, we formulate the problem of the aerial STAR-RIS-assisted MEC system with the goal of minimizing the energy consumption of both IoT devices and aerial STAR-RIS. Considering the constrained power and time budget, the formulated problem can be expressed as
\begin{mini!}|b|[1]
	{\boldsymbol{\lambda}, \boldsymbol{q}, \boldsymbol{\phi}, \boldsymbol{\beta}, \boldsymbol{p}} {E(\boldsymbol{\lambda}, \boldsymbol{q}, \boldsymbol{\phi}, \boldsymbol{\beta}, \boldsymbol{p})}
	{\label{OF}}{}
	\addConstraint{\resizebox{.99\hsize}{!}{$(1-\lambda_k(n))t_k^{\mathrm{loc}}(n) + \lambda_k(n) t_k^{\mathrm{off}}(n) \leq T_k^{\max}(n), \forall{k} \in \mathcal{K}, \forall{n} \in \mathcal{N}$}}\label{c1}
    \addConstraint{\sum_{k=1}^{K} \lambda_k(n) I_k(n) G_k \leq F^{\max}, \forall{n} \in \mathcal{N}}\label{c2}
    \addConstraint{0 \leq \phi^t_m(n), \phi^r_m(n) < 2\pi, \forall{m} \in \mathcal{M}, \forall{n} \in \mathcal{N}}\label{c3}
	\addConstraint{\beta^r_m(n) + \beta^t_m(n) = 1, \forall{m} \in \mathcal{M}, \forall{n} \in \mathcal{N}}\label{c4}
    \addConstraint{0 \leq \lambda_k(n) \leq 1, \forall{k} \in \mathcal{K}, \forall{n} \in \mathcal{N}}\label{c5}
    \addConstraint{0 \leq p_k(n) \leq p_k^{\max}(n), \forall{k} \in \mathcal{K}, \forall{n} \in \mathcal{N}}\label{c6}
    \addConstraint{\frac{||\boldsymbol{q}(n+1)-\boldsymbol{q}(n)||}{I} \leq V^{\max}, \forall{n} \in \mathcal{N}}\label{c7}
    \addConstraint{\boldsymbol{q}(1) = \boldsymbol{q}(N),}\label{c8}
\end{mini!}
where constraint (\ref{c1}) represents the maximum tolerable latency for completion of the task. The MEC server's computational resources is indicated by constraint (\ref{c2}). Constraints (\ref{c3}) and (\ref{c4}) are feasible phase shift values and amplitudes for the coefficients of transmission and reflection of aerial STAR-RIS. The task offloading portion is represented by constraint (\ref{c5}), while the power budget is expressed by constraint (\ref{c6}). Finally, the velocity for aerial STAR-RIS is constrained by (\ref{c7}), and constraint (\ref{c8}) ensures that the aerial STAR-RIS will return to its starting point at the end of the flight duration. The problem is non-convex owing to couplings in both the objective function and constraints, and contains non-linear constraints. Therefore, solving problem (\ref{OF}) within a polynomial time frame conveys a significant challenge. Furthermore, given the dynamic nature of the environment and the potential for unpredictable consequences, DRL emerges as a suitable approach for addressing these challenges. This is primarily attributed to the capability of DRL to effectively address complicated optimization problems that involve high-dimensional spaces. The utilization of DRL in our proposed system has the potential to enhance performance in comparison with traditional methods due to its ability to learn from interactions and make real-time decisions with regards to environmental changes. We employ PPO as one of the DRL techniques due to its sample efficiency, stability, and capability to handle both discrete and continuous actions.

\vspace{-0.2in}
\section{Solution Approach}
In order to implement DRL, we first need to define MDP which serves as a fundamental framework to model and solve sequential decision-making problems in a stochastic environment. The components of MDP include state space $\mathcal{S}$, action space $\mathcal{A}$, reward $\mathcal{R}$, and discount factor $\mathcal{\tau}$.
\subsubsection{State space $\mathcal{S}$} Each state $s_t \in \mathcal{S}$ at time $t$ can be defined as a tuple of $ s_t = \{ \mathbf{h}_{M,B}(n),$ $\mathbf{h}_{r,M}(n), \mathbf{h}_{t,M}(n), \Theta^{\delta}(n), \boldsymbol{q}(n), I_k(n), G_k, T_k^{\max}(n),$ $\delta \in \{r, t\}, \forall{k} \in \mathcal{K}, \forall{n} \in \mathcal{N} \} $.
\subsubsection{Action space $\mathcal{A}$} Each action $a_t \in \mathcal{A}$ at time $t$ contains offloading portion $\lambda_k$, position of aerial STAR-RIS $\boldsymbol{q}$, phase shift coefficient vector $\boldsymbol{\phi}$, amplitude vector $\boldsymbol{\beta}$, and transmit power $\boldsymbol{p}$. Every action for our system is uniquely defined. The position of aerial STAR-RIS, phase shift coefficient vector, amplitude, and transmit power are defined as the incremental value of the current value, and can be expressed as
\begin{equation}
    \boldsymbol{q}^{t+1} = \boldsymbol{q}^{t} \odot \triangle \boldsymbol{q}^{t},
\end{equation}
\begin{equation}
    \boldsymbol{\beta}^{t+1} = \boldsymbol{\beta}^{t} \odot \triangle \boldsymbol{\beta}^{t},
\end{equation}
\begin{equation}
    \boldsymbol{\phi}^{t+1} = \boldsymbol{\phi}^{t} \odot \triangle \boldsymbol{\phi}^{t},
\end{equation}
\begin{equation}
    \boldsymbol{p}^{t+1} = \boldsymbol{p}^{t} \odot \triangle \boldsymbol{p}^{t},
\end{equation}
where $\odot$ is the Hadamard product, and $\triangle$ represents the incremental value. The offloading portion is defined as $\boldsymbol{\lambda} \in (0,1)$.
\subsubsection{Reward $\mathcal{R}$}In order to minimize the total energy consumption, while guaranteeing the conditions of tolerable latency, computational resources, and velocity must be satisfied. Consequently, the reward can be defined as follows.
\begin{align}
&\begin{aligned}
    \mathcal{R} = &\mu_1 E(\boldsymbol{\lambda}, \boldsymbol{q}, \boldsymbol{\phi}, \boldsymbol{\beta}, \boldsymbol{p}) + \sum_{k=1}^{K} \mu_2 C(T_k^{\max}(n) - (1-\lambda_k(n)) \\
    &t_k^{\mathrm{loc}}(n)+\lambda_k(n) t_k^{\mathrm{off}}(n))+ \sum_{k=1}^{K} \mu_3 C(F^{\max} - \lambda_k(n) I_k(n) \\
    &G_k)+\mu_4 C(V^{\max}I - ||\boldsymbol{q}(n+1)-\boldsymbol{q}(n)||), \\
\end{aligned}
\end{align}
where $\mu_1, \mu_2, \mu_3$ and $\mu_4$ are the weight coefficients, and $C(x)$ is the piece-wise function, which is defined as
\begin{equation}
    C(x) = \left \{ \begin{array}{ll}{G_0} & {\text{if } x \geq 0}, \\ {x,} & {\text {otherwise,}}\end{array}\right.
\end{equation}
where $G_0$ is referred to as a positive constant representing revenue.
\subsubsection{Discount factor $\tau$} The discount factor $\tau$ is a parameter used to regulate the significance of future rewards in the agent's decision-making process. The discount factor influences how the agent values immediate rewards compared to rewards received in the future.

Our algorithm operates by having the agent at the BS get the network state information $s_t$ from the environment. Subsequently, the agent proceeds to observe the channel gain, positions of IoT devices and aerial STAR-RIS, computation tasks information. The agent comprises the policy (actor) network and value (critic) network. The actor model has the stochastic policy model $\pi_{\varphi}$ with learnable parameter $\varphi$, which learns to execute actions based on observations. The policy $\pi_{\varphi}$ takes the current observed states $s_t$ as input and outputs a probability distribution over available actions $a_t$. Under the given policy $\pi$, the true value of an action $a$ at state $s$ can be represented as
\begin{equation}
    Q_\pi(s,a) = \mathbb{E}\left[ \mathcal{R}_1 + \tau \mathcal{R}_2 + \dots |s_t=s, a_t=a \right].
\end{equation}
A value network is also initialized to estimate the expected cumulative reward which can be expressed as follows.
\begin{equation}
    V_\pi(s) = \mathbb{E}_\pi \left[ \sum_{t=0}^{T-1} \tau^t \mathcal{R}_t | s_t=s \right].
\end{equation}
The critic comprises the advantage function which is used to evaluates the quality of actions and guide policy updates. It is defined as follows.
\begin{equation}
    A_t = Q_\pi(s,a) - V_\pi(s).
\end{equation}
Henceforth, the surrogate objective function of PPO is to achieve the policy that maximizes the expected cumulative reward while preserving a trust region to prevent overly aggressive policy updates. This can be described as follows.
\begin{equation}
    \mathcal{L}^{\mathrm{CLIP}}(\varphi) = \mathbb{E}_t\left[ \min \left(\hat{r}_t(\varphi) A_t, \mathrm{clip}(\hat{r}_t(\varphi), 1-\epsilon, 1+\epsilon)A_t \right)\right],
\end{equation}
where $\hat{r}_t(\varphi) = \frac{\pi_{\varphi}(a_t|s_t)}{\pi_{\varphi_\mathrm{old}}(a_t|s_t)}$ is the probability ratio. If $\hat{r}_t(\varphi) > 1$, the action is more conceivable than it was in the previous policy, and $0 < \hat{r}_t(\varphi) < 1$ means it is less conceivable. $\epsilon$ is the clipping parameter, and the clipping mechanism serves to ensure the stability of the training process by preventing massive policy changes that might cause training divergence. The overall algorithm is explained in Algorithm \ref{algo1}.
\begin{algorithm}[t] \caption{Aerial STAR-RIS-assisted MEC system}\label{algo1}
	\begin{algorithmic}[1]
	    \For{iteration=$1,2,\dots$}
	        \For{actor=$1,2,\dots$}
	            \State Collect states, actions, reward from a set of tra-
                \NoNumber jectories and implement ${\pi_\theta}_{\mathrm{old}}$ for $T$ in the network
                \NoNumber environment.
	            \State Compute the advantage estimates $A_1, \dots, {A}_{T}$
	        \EndFor
        \State Update the policy by optimizing $\mathcal{L}^{\mathrm{CLIP}}(\varphi)$
        \State Update $\varphi_\mathrm{old} \leftarrow \varphi$
	    \EndFor
	\end{algorithmic}
\end{algorithm}

\begin{figure*}[t!]
	\centering
	\subfigure[]{\includegraphics[width=0.32\linewidth]{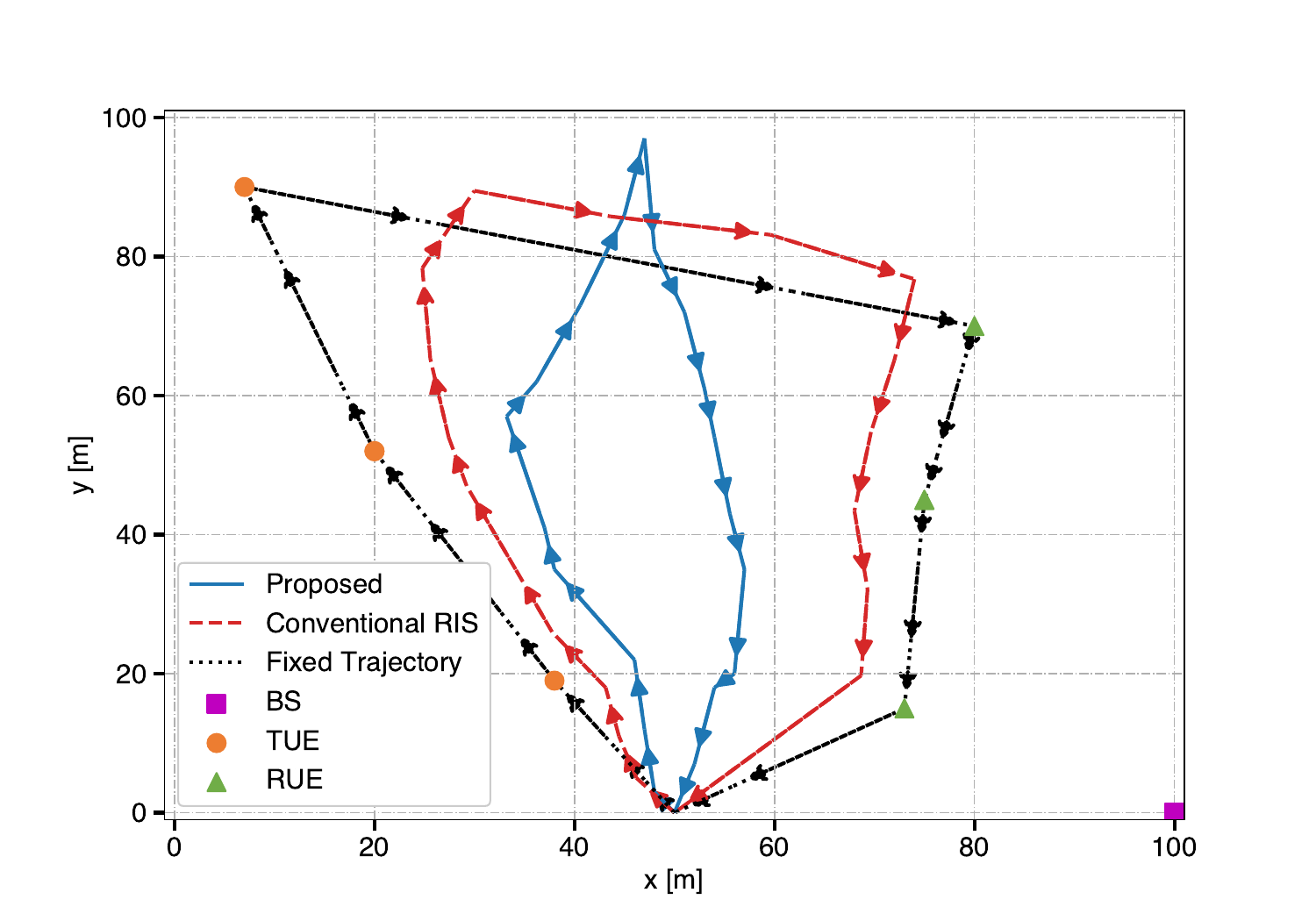}\label{trajectory}}
	\hfil
	\centering
	\subfigure[]{\includegraphics[width=0.32\linewidth]{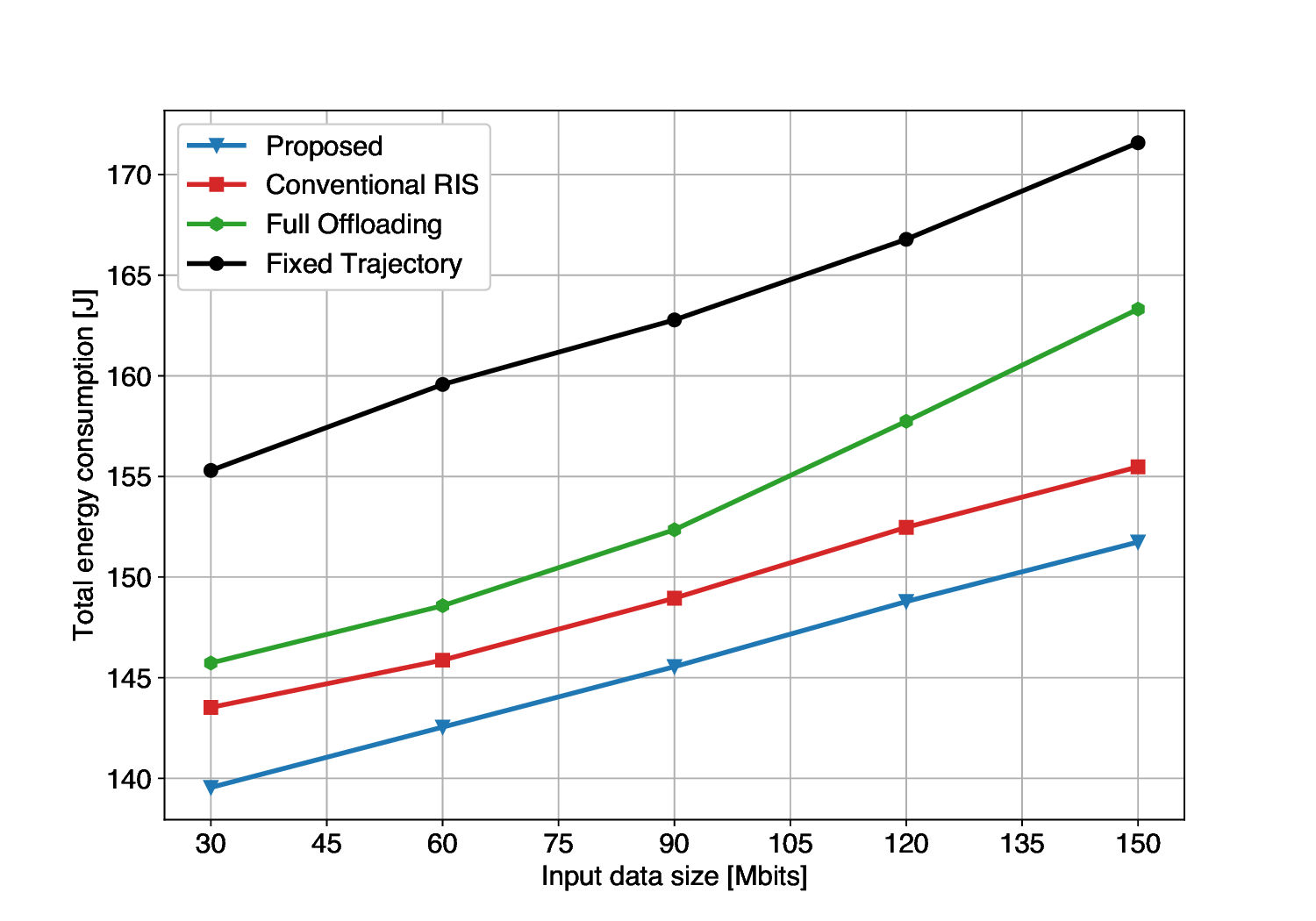}\label{ecvsinput}}
	\hfil
	\centering
 	\subfigure[]{\includegraphics[width=0.32\linewidth]{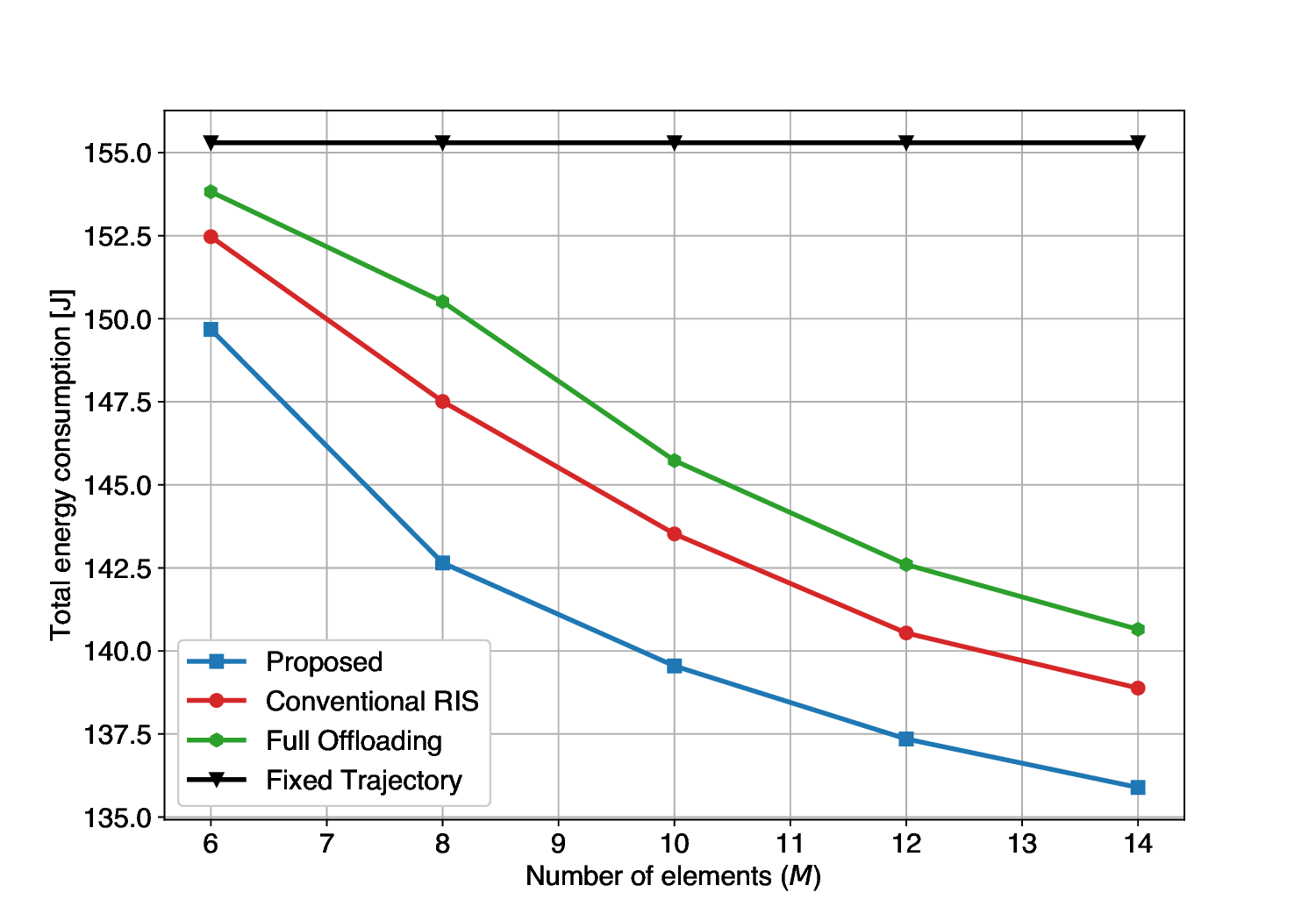}\label{ecvselements}}
	\hfil
	\caption{(a) Trajectories of different algorithms. (b) Performance comparison of total energy consumption under different input data size. (c) Performance comparison of total energy consumption under different number of elements.}
    \label{Metric}
\end{figure*}

\vspace{-0.2in}
\section{Performance Evaluation}
In this section, we evaluate our proposed system to determine its effectiveness through numerical results. The network architecture comprises a square area of 100 m $\times$ 100 m, where a single aerial STAR-RIS initiates and terminates its trajectory at the same coordinate, $\boldsymbol{q}(1) = \boldsymbol{q}(N)= (50,0)$, while hovering at a 20 m altitude. The IoT devices are uniformly distributed, with three IoT devices positioned in the transmission region and three in the reflection region. The simulation parameters are determined as $V^{\max}=$ 7 m/s, $\ddot{R}=$ 10, $T_k^{\max}(n)=[$1, 5$]$ s, $\sigma^2=$ -174 dBm/Hz, $W=$ 10 MHz, $G_k=$ 800 cycles/bits, $f_k=$ 100 MHz, $I_k=[$0.05, 0.1$]$ Mbits, respectively \cite{zhang2023resource, liu2020resource, aung2023energy}. For the benchmark schemes, we compare our results with 1) fixed trajectory, where the UAV serves each IoT device by following a predetermined shortest route with predefined values for $\boldsymbol{\lambda}, \boldsymbol{\phi}, \boldsymbol{\beta}, \boldsymbol{p}$, and 2) conventional RIS, where the RIS is replaced with STAR-RIS and execute PPO to solve the problem. In the latter scenario, the elements only enable reflection ($\beta_r = 1, \beta_t = 0$) when it's in the reflected region and vice versa. The total flight time, $T =$ 50 s, is equally divided into 20 slots.

According to Fig. \ref{trajectory}, we can find that the trajectory of the conventional RIS system has to travel closer to the IoT devices to effectively offload the input data while managing the power budget. In our proposed system, the trajectory is minimal due to the elements supporting both reflection and transmission, enabling the provision of services to IoT devices in both regions. In the fixed trajectory, the UAV is required to navigate toward each IoT device, resulting in greater energy consumption. In Fig. \ref{ecvsinput}, we illustrate the total energy consumption with the input data size. In the full offloading scheme, the IoT device transmits all the computation tasks, i.e., ($\lambda_k(n)=1$) to the MEC server through the aerial STAR-RIS. We can observe that the overall energy consumption of all algorithms grows as the amount of input data increases. Among them, our aerial STAR-RIS-assisted MEC system achieves the best performance because it can adaptively modify the values of amplitudes and phase shift coefficients. Furthermore, Fig. \ref{ecvselements} demonstrates how the number of elements can affect the performance of total energy consumption. With the increase in the number of elements, there is a degradation in energy consumption. This implies that installing a large amount of elements is not necessary for the small network with 6 IoT devices. Among them, our proposed aerial STAR-RIS-assisted MEC system achieves the best performance due to providing the flexibility of energy splitting between the reflection and transmission of each element.

\vspace{-0.2in}
\section{Conclusion}
In this letter, we have investigated the joint optimization of task offloading, aerial STAR-RIS trajectory, amplitude and phase shift coefficients of elements, and transmit power within the constrained power and time budget for the energy-efficient aerial STAR-RIS-assisted MEC system. Given the non-convex structure of the problem and the dynamic nature of the environment, we proposed PPO as a solution approach due to its ability to handle high dimensional spaces, sample efficiency, and stability. To prove the efficacy of our proposed system, we compared our results with several benchmark schemes, including the conventional RIS-assisted system and the full offloading scenario. The numerical results indicate the superiority of our proposed system in comparison to the benchmark schemes.

\bibliographystyle{IEEEtran}
\vspace{-0.2in}
\bibliography{mybib}

\end{document}